\documentclass[prd,aps,preprint,nofootinbib,superscriptaddress]{revtex4}

\usepackage{amsmath}
\usepackage{epsfig}
\usepackage{graphicx}
\usepackage{amssymb}
\usepackage{subfigure}

\def\beqn{\begin{eqnarray}} 
\def\eeqn{\end{eqnarray}} 
\def\be{\begin{equation}}
\def\ee{\end{equation}}

\newcommand{\msun}{M_{\odot}{\ }}

\begin{document}


\title{Gamma Rays from Cosmic-Ray Proton Scattering in AGN Jets:\\ the Intra-Cluster Gas vastly outshines Dark Matter}

\author{Stefano Profumo}
\email{profumo@scipp.ucsc.edu} \affiliation{Department of Physics, University of California, 1156 High St., Santa Cruz, CA 95064, USA}\affiliation{Santa Cruz Institute for Particle Physics, Santa Cruz, CA 95064, USA} 
\author{Lorenzo Ubaldi}
\email{ubaldi@th.physik.uni-bonn.de}
\affiliation{Bethe Center for Theoretical Physics and Physikalisches Institut, Universit\"at Bonn, Bonn, Germany}
\author{Mikhail Gorchtein}
\email{gorshtey@kph.uni-mainz.de}
\affiliation{%
Institut f\"ur Kernphysik, Universit\"at Mainz
55128 Mainz, Germany
}
\date{\today}

\begin{abstract}
\noindent Active Galactic Nuclei (AGN) host powerful jets containing high-energy electrons and protons. The astrophysical environment where AGNs and their jets are found is characterized by large concentrations of both dark matter (DM) and intra-cluster medium (ICM) gas. As the high-energy jet particles transverse the DM and the ICM, elastic and inelastic scattering processes generically lead to the production of final-state photons. As first envisioned by Bloom and Wells (1998), and as more recently pointed out by us and others, the scattering of electrons off of DM could lead to a potentially detectable gamma-ray signal, with the parton-level contribution from protons offering dimmer perspectives. Recently, Chang et al. argued that taking into account photons from hadronization and showering, the actual photon flux is substantially increased. Here, we point out that the proton-jets have to be highly collimated, contrary to what predicted by simple blob-geometry jet-models sometimes employed in these studies, otherwise they would produce a very large flux of photons from inelastic collisions with ICM nucleons, which would outshine by many orders of magnitude the signal from DM, for almost any reasonable ICM and DM density profiles.
\end{abstract}

\maketitle

Active Galactic Nuclei (AGN) are sources of powerful collimated jets containing high-energy particles possibly including electrons and protons, and are believed to exist in environments typically hosting large densities of Dark Matter (DM), such as the center of massive elliptical galaxies. Ref.~\cite{Bloom:1997vm} pointed out a long time ago that electron-DM scattering near AGNs could produce multi-GeV gamma rays, estimated the process from a particle physics standpoint and the overall expected photon rate, and concluded that the signal was too suppressed to be detectable by current or planned gamma-ray telescopes.

More recently, Ref.~\cite{Gorchtein:2010xa} and \cite{Huang:2011dg} have elaborated on the original idea proposed by \citet{Bloom:1997vm}, and have evaluated in detail the scattering of relativistic electrons in AGN jets off of DM, arguing that indeed the process can produce a gamma-ray signal potentially detectable by the Fermi Large Area Telescope (LAT). From a theoretical and observational standpoint, AGN jets can be either of predominantly leptonic or hadronic nature, or they can contain both electrons and protons \cite{Dermer} . In the case of hadronic jets, the scattering off of DM particles happens at the parton level, and the resulting gamma-ray signal is a few orders of magnitude weaker compared to the electron case, after integrating on the parton distribution function (PDF). The analyses of both Ref.~\cite{Gorchtein:2010xa} and \cite{Huang:2011dg}, however, only included the parton-level events, and did not include the effects of showering and hadronization. 

In Ref.~\cite{Gorchtein:2010xa} we argued, based on quantitative estimates, that it appeared unlikely to get any significant enhancement from photons produced in the showering of proton remnants. Recently, however, Chang et al. reached a different conclusion~\cite{Chang:2012sk}, computing the process in detail and including hadronization, and claimed that the resulting gamma ray signal can be significantly affected and increased, under certain assumptions on the AGN jet structure. The photons from proton-DM collisions are classified as 
\begin{enumerate}
\item Final State Radiation (FSR). They dominate at large scattering angles and for photon energies up to $\delta M$, typically between a few GeV and $\sim 100$ GeV, with $\delta M$ the mass difference between the mediator\footnote{In supersymmetric models, for example, the DM would be the neutralino and the mediator a squark.} of the scattering process and the DM. Compared to the parton-level process, these gamma rays are enhanced by hadronization due to large multiplicity and lack of suppression by the QED fine structure constant.
\item Shower from proton remnants.  They are emitted along the proton's incoming direction and would affect significantly the spectrum at higher photon energies ($E_\gamma > \delta M$). One therefore needs a large enough flux of protons {\em in the direction of our line of sight} for this contribution to be relevant.
\end{enumerate}

In this short note we point out that if there is a significant flux of protons in the jet pointed along our line of sight, the photons produced by inelastic collisions of the jet protons with nuclei in the intra-cluster medium\footnote{While the ICM consists primarily of ionized hydrogen and helium, we shall assume here for simplicity that it be composed of ionized hydrogen (i.e. protons) only.} (ICM) would {\em by far outshine} those produced by scattering off of dark matter. We provide here both simple estimates and a detailed worked-out example for the case of the nearby AGN Centaurus A (CenA) to substantiate our claim.


Let us start with a simple estimate, and consider the two processes of proton-DM and of proton-proton scattering. The photon flux from the two scattering processes can be cast, respectively, as
\beqn \label{master}
\left(\frac{d\phi_\gamma}{dE_\gamma}\right)_B &=& \frac{\delta_B}{d_{AGN}^2 m_p} \int \ dE_p \frac{d\sigma_{pp}(E_p)}{dE_\gamma} \left.\frac{d\phi_p}{d\Omega_p dE_p} \right|_{\theta = \theta_{AGN}} , \label{eq:fluxB} \\
\left(\frac{d\phi_\gamma}{dE_\gamma}\right)_{DM} &=& \frac{\delta_{DM}}{d_{AGN}^2 m_{DM}} \int \ dE_p \frac{d\sigma_{pDM}(E_p)}{dE_\gamma} \left.\frac{d\phi_p}{d\Omega_p dE_p} \right|_{\theta = \theta_{AGN}} , \label{eq:fluxDM} 
\eeqn
where the label $B$ in the first line stands for ``baryons'', while $DM$ in the second line stands for dark matter. $d_{AGN}$ is the distance to the AGN ($d_{AGN} = 3.7$ Mpc for the case of CenA), $\delta_{B(DM)} = \int \ \rho_{B(DM)}(r) \ dr$, with $\rho_{B(DM)}(r)$ the baryonic (DM) density profile (the factor $1/m_{B,\ DM}$ turns this into the actual number density). $\frac{d\sigma_{pp}(E_p)}{dE_\gamma}$ indicates the differential cross section for the process $pp \to \pi_0 + X \to \gamma\gamma + X$, while $\frac{d\sigma_{pDM}(E_p)}{dE_\gamma}$ is for the process $p+DM \to \gamma + \dots$. The last term in the integral, $\left. \frac{d\phi_p}{d\Omega_p dE_p} \right|_{\theta = \theta_{AGN}}$, is the energy distribution of the protons in the jet, in the direction of our line of sight (for CenA, $\theta_{AGN}= 68^\circ$). We will assume here a simple blob geometry to describe the jet, which implies that the jet is not perfectly collimated, but that there are some high energy protons that point away from the jet axis. Note that Refs.~~\cite{Gorchtein:2010xa} and \cite{Huang:2011dg}, while assuming the blob geometry, impose by hand a cutoff on the angle of the protons in the black hole frame in order to obtain a collimated jet. The authors of Ref.~\cite{Chang:2012sk}, instead, contemplate both a cut-off and a full blob geometry. The latter includes ``line-of-sight'' protons which are traveling towards us and that would be responsible for most of the gamma-ray signal from proton-DM scattering at energies higher than $\delta M$, as we mentioned above.


Let us first compare the relevant differential photon fluxes for the two processes (i.e. from scattering of high-energy cosmic-ray protons from the jet off of ICM protons versus off of dark matter). The overall factors appearing in Eq.~(\ref{master}) for proton-DM versus proton-ICM proton differ by three elements: 

(i) for proton-DM scattering, the total cross section is at most $10^{-5}$ mb~\cite{Chang:2012sk}, while for the inelastic process $pp \to \pi_0+X$, the inclusive inelastic cross section is about 30 mb, i.e. {\em six orders of magnitude larger}.

(ii) since the $\gamma$-ray flux depends on the number density of targets, and not on the physical density, the fluxes are inversely proportional to the mass of the target particles, which for DM masses of order 100 GeV, as contemplated in these studies, gives an {\em extra suppression of  two orders of magnitude} for proton-DM compared to proton-proton.

(iii) the last, critical element is the comparison of the DM and baryon density profiles. While there exist large observational uncertainties, numerical simulations indicate that the physical densities of baryons and DM are, over the relevant radii, comparable\footnote{See for example figure 4 of~\cite{Fedeli:2011gj}. To test this in the innermost regions (where the jet structure becomes, however, increasingly uncertain)  one would have to extrapolate to much smaller radii than what showed e.g. in Ref.~\cite{Fedeli:2011gj}, but the trend indicates that the density of cold gas is generally higher than that of DM as we approach the center}. Assuming that is the case, the values of $\delta_B$ and $\delta_{DM}$ are of the same order. We will quantitatively test caveats to this assumption (which might depend on a steep density profile for $r\to0$) below.

Thus, from this preliminary estimate, we expect the photon flux from proton-proton scattering to be several orders of magnitude (between 6 and 8, but possibly more) larger than the one from proton-DM. If indeed one allowed a jet geometry with a fraction of off-axis protons, as in the naive blob model without a cutoff, one would conclude that the photon flux from proton-ICM proton inelastic scattering would completely 
outshine the process involving dark matter and violate gamma-ray observations by many orders of magnitude.

We intend to test here the robustness of the conclusion we reached in the qualitative discussion above, and assess possible caveats to that conclusion by contemplating the possibility that a very steep DM density profile compensate for the many orders of magnitude larger $\gamma$-ray flux expected from scattering off of gas nuclei. We parametrize the density profiles for the baryonic and dark matter target densities for small radii $r\to0$ by the following simplified functional forms: 
\begin{eqnarray}
\rho_{DM}&\simeq&\rho_{0,DM}\left(\frac{r}{r_c}\right)^{-\alpha},\\
\rho_{B}&\simeq&\rho_{0,B}\left(\frac{r}{r_c}\right)^{-\beta}.
\end{eqnarray}
While we assume the same scaling radius, $\delta_{B,DM}$ will simply scale linearly with an overall normalization change induced by a change in $r_c$ for either one of the two species, so taking this effect into account is trivial. The key aspect is what happens when $\alpha$ and $\beta$ are drastically different.

We normalize the overall factors $\rho_0\ r_c^{\alpha,\beta}$ to be of the same order of magnitude as the mass enclosed within 5 kpc from the AGN: in the case of CenA the total gravitating mass enclosed in 5 kpc is equal to $7.5\times 10^{10} \ \msun$, as obtained from X-ray measurements under the hypothesis of hydrostatic equilibrium~\cite{Kraft:2003gp}. Notice that this is, for our purposes, a conservative assumption, as the DM density is generally thought to be subdominant (if not negligible!) compared to the baryonic mass in the vicinity of the AGN center (see e.g. \cite{Mamon:2004sz}). To compute the factor $\delta_B$ in Eq.~(\ref{master}), we set $r_{min}=10 \ R_S$, with $R_S = 5\times 10^{-6}$ pc the Schwarzschild radius for CenA, and $r_{max} = 15$ kpc. The integral has a very mild dependence on $r_{min}$ when the latter is chosen between a few and 1000 $R_S$, while it is very insensitive on the upper limit $r_{max}$ (which depends on how far the jet extends from the AGN).

\begin{figure}[t]
\begin{center}$
\begin{array}{cc}
\includegraphics[width=0.45\textwidth]{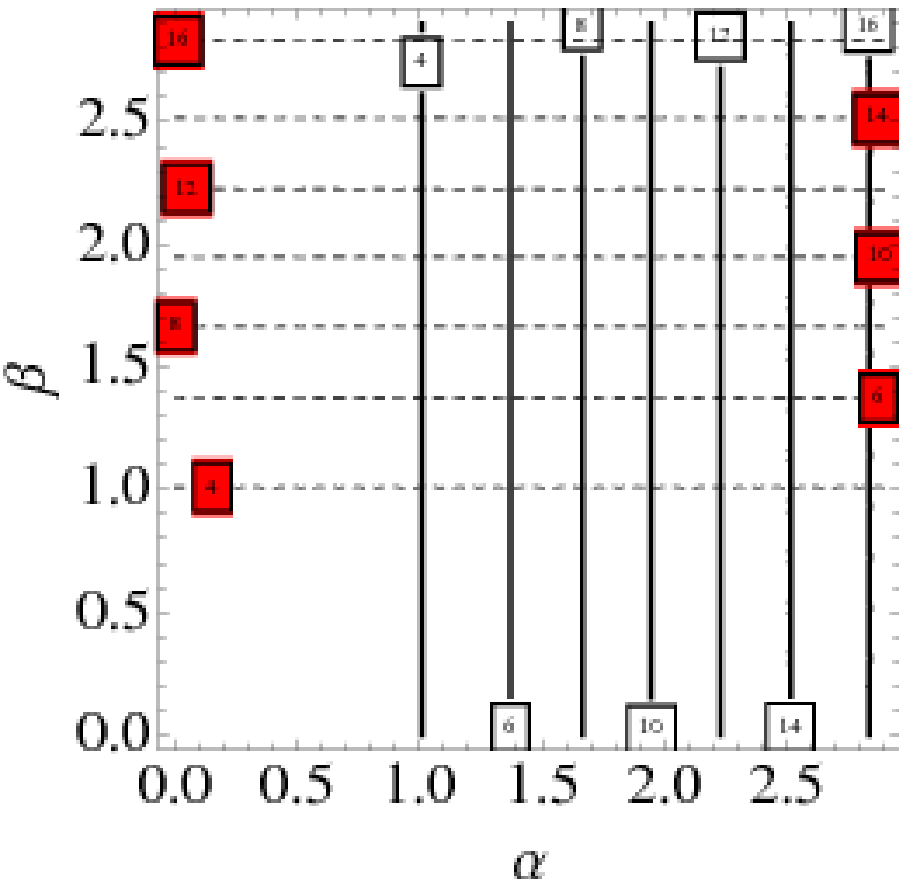} &
\includegraphics[width=0.45\textwidth]{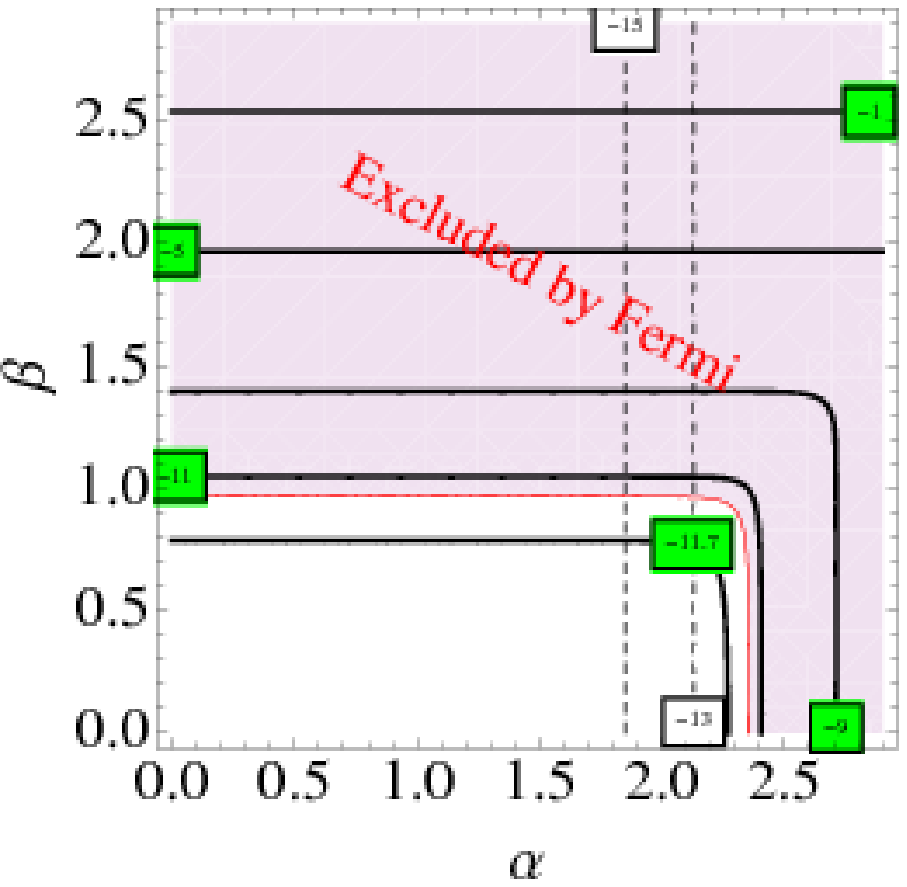}
\end{array}$
\end{center}
\caption{In these figures $\alpha$ and $\beta$ are the slopes of the DM and baryon profiles, respectively. In the plot on the left the vertical lines are contours of $\log_{10}(\delta_{DM})$, while the horizontal lines are contours of $\log_{10}(\delta_{B})$. On the right we show contour lines (thick solid) of $\log_{10}\left[ E_\gamma^2 \left(\frac{d\phi_\gamma}{dE_\gamma}\right)_B  + E_\gamma^2 \left(\frac{d\phi_\gamma}{dE_\gamma}\right)_{DM}  \right]$, for $E_\gamma=10$ GeV. The two vertical dashed lines show that the flux only from DM, $\log_{10}\left[  E_\gamma^2 \left(\frac{d\phi_\gamma}{dE_\gamma}\right)_{DM}  \right]$, can be below the observational limits for lower values of the slope $\alpha$. The red line corresponds to a flux of $5\times 10^{-12}$ erg s$^{-1}$ cm$^{-2}$, value measured by Fermi at 10 GeV, and the shaded region above it is then excluded. }
\label{figure}
\end{figure}

In the left panel of Fig.~\ref{figure} we vary the steepness of the density profiles $\alpha$ and $\beta$ for the dark matter ($\alpha$) and for the baryonic ($\beta$) densities, and show curves of constant $\delta_{B, DM}$. The labels indicate curves at the corresponding values of  $\log_{10}(\delta_{DM})$ (white labels) and $\log_{10}(\delta_{B})$ (red labels). The contours saturate at close to constant values towards the lower left corner. As we anticipated in the qualitative discussion above, for comparable steepness $\alpha,\beta$, we in fact find $\delta_{DM}\sim\delta_B$.

We now proceed to the calculation of the $\gamma$-ray flux from jet protons scattering off of DM versus ICM protons. For the DM calculation, we closely follow the assumptions in Ref.~\cite{Chang:2012sk}, and we employ here the following fiducial values for the dark matter particle properties: a mass ($m_{DM}$) of 300 GeV, and a total cross section ($\sigma_{pDM}$) of $10^{-5}$ mb. 

To calculate the photon flux from proton-proton interactions, the relevant differential cross section can be written as~\cite{Dermer}
\be \label{eq:crosspp}
\frac{d\sigma_{pp}(E_p)}{dE_\gamma} \simeq 2\sigma_{pp,\pi_0}(E_p) \delta(E_\gamma - \chi E_p),
\ee
where $\sigma_{pp,\pi_0} \sim 30$ mbarn is the inclusive inelastic cross section for the process $pp \to \pi_0$ for energies $E_p>10$ GeV. The delta function is an approximation for gamma rays formed in the reaction $pp\to \pi_0 \to \gamma\gamma$, with $\chi \sim 0.05$ the ratio of the mean energy of the produced photon to the energy of the incident proton.

Assuming a simple blob geometry to describe the cosmic-ray proton jet, and again employing the same assumptions as in Ref.~\cite{Chang:2012sk}, we write the proton energy distribution in the black hole frame as
 \beqn
\frac{d\phi_p}{d\Omega_p dE_p} & = & \frac{k_p}{4\pi} \left( \frac{E}{E_0} \right)^{-2} \left[ \Gamma_B (1-\beta_B \cos\theta) \right]^{-3}, \\
k_p & = & \frac{L_p}{E_0^2 \Gamma_B \ln(E_{max}/E_{min})},
\eeqn
where $\Gamma_B = 3$ is the bulk Lorentz factor (of the blob with respect to the black hole), which corresponds to a velocity $\beta_B = 0.94$, $L_p = 10^{45}$ erg/s is the luminosity of the jet, $E_{min}=10$ GeV and $E_{max} = 10^7$ GeV are the minimum and maximum energies of the protons in the distribution. $E_0$ is some reference energy that does not need to be specified, as it cancels out in the definition above. As a result, we can cast Eq.~(\ref{master}) in the following form:
\be
 \frac{d\phi_\gamma}{dE_\gamma} =  \frac{\delta_B}{d_{AGN}^2 m_p} 2 \sigma_{pp,\pi_0}(E_p)\frac{k_p}{4\pi} \frac{E_0^2}{\chi} \frac{1}{E_\gamma^2} \left[ \Gamma_B (1-\beta_B \cos(68^\circ)) \right]^{-3}.
\ee
We note that the quantity typically shown is the spectral energy distribution, i.e. $\nu S_\nu \equiv E_\gamma^2  \frac{d\phi_\gamma}{dE_\gamma}$: here, $\nu S_\nu$ is close to a constant, as the powers of $E_\gamma$ cancel out. 

The right panel of Fig.~\ref{figure} illustrates curves of constant $\gamma$-ray flux at an energy $E_\gamma=10$ GeV. We shade the region ruled out by Fermi observations of CenA. We note that the entire plot features fluxes in excess of $10^{-12}\ {\rm erg}\ {\rm s}^{-1}\ {\rm cm}^{-2}$ due to the proton-proton contribution. As can be appreciated from the dashed vertical lines, for the DM contribution to overcome the smallest possible proton-proton background one would need DM density profiles much steeper than $\alpha=2$, a value that is extraordinarily hard to achieve in realistic situations, especially in the (most relevant) innermost regions: there, tidal disruption due to stars would very likely make an extremely steep density profile highly unstable. Also, the very large densities achieved at small radii for steep profiles with $\alpha\gtrsim2$ would likely be cutoff by DM pair annihilation \cite{Gondolo:1999ef} at a radius potentially much larger than the value of $R_{min}$ than the one we assumed to calculate $\delta_{DM}$ here.

Besides the steepness of the density profiles, an additional assumption one might consider changing is the picture for the jet structure and geometry. Remember that with the blob geometry, although most of the protons move within a cone along the direction of the jet-axis, there is still a small fraction pointing at large angles with respect to the axis. This small fraction leads to the catastrophic result just described, so one might wish to suppress it. Indeed, observations indicate that AGN jets appear to be very collimated. One way to suppress off-jet protons, is to introduce ``by hand'' a cut-off on the angle in order to restrict the particles to move in a narrow cone centered on the axis, as was done e.g. in~\cite{Gorchtein:2010xa}. In Figure 7 of Ref.~\cite{Chang:2012sk} the authors show results with such a cutoff (orange lines) and without it (black line). In light of our considerations, we argue that the latter should be discarded. 

As a further remark, we wish to point out that if the jet is well collimated, photons from proton remnants should be negligible. Let us consider the production of energetic $\pi^0$'s at large angles, e.g. $\theta\approx68^\circ$, in the case of a collimated proton jet scattering off ICM protons. This process requires one active quark to be bent at a large angle and it involves a 4-momentum transfer in an elementary collision of order $t\sim-E^2$, where $E$ denotes the scattered quark energy. The amplitude for such a process would be suppressed as $(\Lambda/E)^2$, $\Lambda\sim1$ GeV being a characteristic hadronic scale; the differential cross section, correspondingly, would be suppressed as  $d\sigma/d\Omega(\theta=68^\circ)\sim d\sigma/d\Omega(\theta=0) (\Lambda/E)^{4}$, implying that for $E_\gamma\sim100$ GeV we would encounter a $10^{-8}$ suppression relative to $\pi^0$ production in the direction of the AGN jet. 

Note that for a predominantly leptonic jet, i.e. the case that Ref.~\cite{Gorchtein:2010xa} and \cite{Huang:2011dg} argued being the most promising one, the relevant associated background of jet electrons scattering off of ICM protons producing bremsstrahlung photons would be by far suppressed compared to the proton-proton background we studied here \cite{Berg:1958zz}. The present discussion, therefore, does not invalidate the results of  Ref.~\cite{Gorchtein:2010xa} and \cite{Huang:2011dg} for a predominantly leptonic jet.

In conclusion, we studied the impact of inelastic proton-proton collisions in AGN jets that could produce a sizable gamma-ray signal from proton-DM scattering. While Ref.~\cite{Chang:2012sk} argued that photons from proton remnant showering could contribute to enhancing the proton-DM signal, particularly at higher photon energies, we showed here that those same jet protons would vastly outshine the proton-DM gamma-ray flux as they inelastically scatter off of nucleons in the intra-cluster medium. We argued that the proton-proton background is at least 6-8 orders of magnitude larger than the DM-proton signal for comparable baryonic and dark matter density profiles. 


Finally, while the process of dark matter scattering off AGN jet particles is potentially observable if the AGN jet is predominantly leptonic (as pointed out in Ref.~\cite{Gorchtein:2010xa} and \cite{Huang:2011dg}), we remain somewhat pessimistic as to the possibility of detecting a signal for the DM-proton case.

\begin{acknowledgments}
\noindent  SP is partly supported by the US Department of Energy and by Contract DE-FG02-04ER41268. LU would like to thank Emilio Romano-Diaz for a helpful discussion on density profiles in AGN. MG is supported by the Deutsche Forschungsgemeinschaft through the Collaborative Research Centre SFB1044.
\end{acknowledgments}


\begin{thebibliography}{10}
\expandafter\ifx\csname natexlab\endcsname\relax\def\natexlab#1{#1}\fi
\expandafter\ifx\csname bibnamefont\endcsname\relax
  \def\bibnamefont#1{#1}\fi
\expandafter\ifx\csname bibfnamefont\endcsname\relax
  \def\bibfnamefont#1{#1}\fi
\expandafter\ifx\csname citenamefont\endcsname\relax
  \def\citenamefont#1{#1}\fi
\expandafter\ifx\csname url\endcsname\relax
  \def\url#1{\texttt{#1}}\fi
\expandafter\ifx\csname urlprefix\endcsname\relax\def\urlprefix{URL }\fi
\providecommand{\bibinfo}[2]{#2}
\providecommand{\eprint}[2][]{\url{#2}}

\bibitem[{\citenamefont{Bloom and Wells}(1998)}]{Bloom:1997vm}
\bibinfo{author}{\bibfnamefont{E.~D.} \bibnamefont{Bloom}} \bibnamefont{and}
  \bibinfo{author}{\bibfnamefont{J.~D.} \bibnamefont{Wells}},
  \bibinfo{journal}{Phys.Rev.} \textbf{\bibinfo{volume}{D57}},
  \bibinfo{pages}{1299} (\bibinfo{year}{1998}), \eprint{astro-ph/9706085}.

\bibitem[{\citenamefont{Gorchtein et~al.}(2010)\citenamefont{Gorchtein,
  Profumo, and Ubaldi}}]{Gorchtein:2010xa}
\bibinfo{author}{\bibfnamefont{M.}~\bibnamefont{Gorchtein}},
  \bibinfo{author}{\bibfnamefont{S.}~\bibnamefont{Profumo}}, \bibnamefont{and}
  \bibinfo{author}{\bibfnamefont{L.}~\bibnamefont{Ubaldi}},
  \bibinfo{journal}{Phys.Rev.} \textbf{\bibinfo{volume}{D82}},
  \bibinfo{pages}{083514} (\bibinfo{year}{2010}), \eprint{1008.2230}.

\bibitem[{\citenamefont{Huang et~al.}(2012)\citenamefont{Huang, Rajaraman, and
  Tait}}]{Huang:2011dg}
\bibinfo{author}{\bibfnamefont{J.-R.} \bibnamefont{Huang}},
  \bibinfo{author}{\bibfnamefont{A.}~\bibnamefont{Rajaraman}},
  \bibnamefont{and} \bibinfo{author}{\bibfnamefont{T.~M.} \bibnamefont{Tait}},
  \bibinfo{journal}{JCAP} \textbf{\bibinfo{volume}{1205}}, \bibinfo{pages}{027}
  (\bibinfo{year}{2012}), \eprint{1109.2587}.

\bibitem[{\citenamefont{{Dermer} and {Menon}}(2009)}]{Dermer}
\bibinfo{author}{\bibfnamefont{C.~D.} \bibnamefont{{Dermer}}} \bibnamefont{and}
  \bibinfo{author}{\bibfnamefont{G.}~\bibnamefont{{Menon}}},
  \emph{\bibinfo{title}{{High Energy Radiation from Black Holes: Gamma Rays,
  Cosmic Rays, and Neutrinos}}} (\bibinfo{year}{2009}).

\bibitem[{\citenamefont{Chang et~al.}(2012)\citenamefont{Chang, Gao, and
  Spannowsky}}]{Chang:2012sk}
\bibinfo{author}{\bibfnamefont{S.}~\bibnamefont{Chang}},
  \bibinfo{author}{\bibfnamefont{Y.}~\bibnamefont{Gao}}, \bibnamefont{and}
  \bibinfo{author}{\bibfnamefont{M.}~\bibnamefont{Spannowsky}},
  \bibinfo{journal}{JCAP} \textbf{\bibinfo{volume}{1211}}, \bibinfo{pages}{053}
  (\bibinfo{year}{2012}), \eprint{1210.1870}.

\bibitem[{\citenamefont{Fedeli}(2012)}]{Fedeli:2011gj}
\bibinfo{author}{\bibfnamefont{C.}~\bibnamefont{Fedeli}},
  \bibinfo{journal}{Mon.Not.Roy.Astron.Soc.} \textbf{\bibinfo{volume}{424}},
  \bibinfo{pages}{1244} (\bibinfo{year}{2012}), \eprint{1111.5780}.

\bibitem[{\citenamefont{Kraft et~al.}(2003)\citenamefont{Kraft, Vazquez,
  Forman, Jones, Murray et~al.}}]{Kraft:2003gp}
\bibinfo{author}{\bibfnamefont{R.~P.} \bibnamefont{Kraft}},
  \bibinfo{author}{\bibfnamefont{S.}~\bibnamefont{Vazquez}},
  \bibinfo{author}{\bibfnamefont{W.}~\bibnamefont{Forman}},
  \bibinfo{author}{\bibfnamefont{C.}~\bibnamefont{Jones}},
  \bibinfo{author}{\bibfnamefont{S.}~\bibnamefont{Murray}},
  \bibnamefont{et~al.}, \bibinfo{journal}{Astrophys.J.}
  \textbf{\bibinfo{volume}{592}}, \bibinfo{pages}{129} (\bibinfo{year}{2003}),
  \eprint{astro-ph/0304363}.

\bibitem[{\citenamefont{Mamon and Lokas}(2005)}]{Mamon:2004sz}
\bibinfo{author}{\bibfnamefont{G.~A.} \bibnamefont{Mamon}} \bibnamefont{and}
  \bibinfo{author}{\bibfnamefont{E.~L.} \bibnamefont{Lokas}},
  \bibinfo{journal}{Mon.Not.Roy.Astron.Soc.} \textbf{\bibinfo{volume}{362}},
  \bibinfo{pages}{95} (\bibinfo{year}{2005}), \eprint{astro-ph/0405466}.

\bibitem[{\citenamefont{Gondolo and Silk}(1999)}]{Gondolo:1999ef}
\bibinfo{author}{\bibfnamefont{P.}~\bibnamefont{Gondolo}} \bibnamefont{and}
  \bibinfo{author}{\bibfnamefont{J.}~\bibnamefont{Silk}},
  \bibinfo{journal}{Phys.Rev.Lett.} \textbf{\bibinfo{volume}{83}},
  \bibinfo{pages}{1719} (\bibinfo{year}{1999}), \eprint{astro-ph/9906391}.

\bibitem[{\citenamefont{Berg and Lindner}(1958)}]{Berg:1958zz}
\bibinfo{author}{\bibfnamefont{R.}~\bibnamefont{Berg}} \bibnamefont{and}
  \bibinfo{author}{\bibfnamefont{C.}~\bibnamefont{Lindner}},
  \bibinfo{journal}{Phys.Rev.} \textbf{\bibinfo{volume}{112}},
  \bibinfo{pages}{2072} (\bibinfo{year}{1958}).

\end{thebibliography}

\end{document}